\documentclass[aip,jcp,reprint,english,superscriptaddress]{revtex4-1}
\usepackage{amsthm}
\usepackage[scanall]{psfrag}
\usepackage{dsfont}
\usepackage[T1]{fontenc}
\usepackage[latin9]{inputenc}
\usepackage{amsmath}
\usepackage{amssymb}
\usepackage{bm}
\usepackage{graphicx}
\usepackage{xcolor}
\usepackage[colorlinks=true]{hyperref}
\usepackage{url}

\newcommand{\mvec}[1]{{\bm{#1}}}
\newcommand{\overbar}[1]{\mkern 1.5mu\overline{\mkern-1.5mu#1\mkern-1.5mu}\mkern 1.5mu}

\makeatletter
\@ifundefined{textcolor}{}
{%
 \definecolor{BLACK}{gray}{0}
 \definecolor{WHITE}{gray}{1}
 \definecolor{RED}{rgb}{1,0,0}
 \definecolor{GREEN}{rgb}{0,1,0}
 \definecolor{BLUE}{rgb}{0,0,1}
 \definecolor{CYAN}{cmyk}{1,0,0,0}
 \definecolor{MAGENTA}{cmyk}{0,1,0,0}
 \definecolor{YELLOW}{cmyk}{0,0,1,0}
 }
\definecolor{michael}{rgb}{0,.8,.5}

\makeatother

\usepackage{babel}

\begin{document}

\title{Potential Energy Landscape of the Two-Dimensional \texorpdfstring{$XY$}{XY} Model: Higher-Index Stationary Points}

\author{Dhagash Mehta}
\email[]{dbmehta@ncsu.edu}
\affiliation{Department of Chemistry, The University of Cambridge, Lensfield Road, Cambridge CB2 1EW, UK}
\affiliation{Department of Mathematics, North Carolina State University, Raleigh, NC 27695-8205, USA}

\author{Ciaran Hughes}
\email[]{ch558@cam.ac.uk}
\affiliation{The Department of Applied Mathematics and Theoretical Physics, The University of Cambridge, Clarkson Road, Cambridge CB3 0EH, UK }

\author{Michael Kastner}
\email[]{kastner@sun.ac.za}
\affiliation{National Institute for Theoretical Physics (NITheP), Stellenbosch 7600, South Africa}
\affiliation{Institute of Theoretical Physics,  University of Stellenbosch, Stellenbosch 7600, South Africa}

\author{David J. Wales}
\email[]{dw34@cam.ac.uk}
\affiliation{University Chemical Laboratories, Lensfield Road, Cambridge CB2 1EW, UK}

\begin{abstract}
The application of numerical techniques to the study of energy
landscapes of large systems relies on sufficient sampling of the
stationary points. Since the number of stationary points is believed
to grow exponentially with system size, we can only sample a small
fraction. We investigate the interplay between this
restricted sample size and the physical features of the potential energy landscape for the two-dimensional $XY$ model
in the absence of disorder with up to $N=100$ spins. Using an eigenvector-following 
technique, we numerically compute stationary points with a given Hessian
index $I$ for all possible values of $I$. We investigate the number of
stationary points, their energy and index distributions, and other related
quantities, with particular focus on the scaling with $N$. The results are used
to test a number of conjectures and approximate analytic
results for the general properties of energy landscapes.
\end{abstract}

\maketitle

\section{Introduction}

The stationary points of a potential energy function, defined as configurations where the gradient of the potential energy function vanishes, play a crucial role in understanding and describing physical and chemical phenomena. Based on these stationary points, a variety of methods, collectively known as ``potential energy landscape theory'', have attracted a lot of attention, with applications to many-body systems as diverse as metallic clusters, biomolecules, structural glass formers, and coarse-grained models of soft matter.\cite{Wales03,RevModPhys.80.167} In all these examples, the potential energy landscape is a multivariate function defined on a high-dimensional manifold.

In most applications, the potential energy function is nonlinear, and an analytic calculation of the stationary points is therefore extremely difficult, and in most cases impossible. Hence, one has to rely on numerical methods. In the present paper we report the results of a numerical computation of stationary points of the $XY$ model in the absence of disorder.

The $XY$ model is among the simplest lattice spin models amenable to an energy landscape approach. The even simpler Ising model has a discrete configuration space and the notion of a stationary point of the potential energy function is somewhat different. Despite the $XY$ model's simplicity, its potential energy landscape exhibits a plethora of interesting properties, and it has been helpful in understanding general features of potential energy landscapes. We consider $d$-dimensional cubic lattices $\Lambda$ of side length $L$, so that the total number of lattice sites is $N=L^d$. For each lattice site $k\in\Lambda$ we assign a degree of freedom, parameterized by the angular variable $\theta_{k}\in(-\pi,\pi]$. The Hamiltonian of the $XY$ model is defined as
\begin{equation}\label{eq:F_phi}
H=\frac{1}{2}\sum_{k\in\Lambda}\, \sum_{l\in \mathcal{N}(k)}[1- \cos(\theta_k-\theta_{l})],
\end{equation}
where $\mathcal{N}(k)$ denotes the set of nearest-neighbors of lattice site $k$. No kinetic energy term is present in \eqref{eq:F_phi}, and the potential energy function is therefore identical to the Hamiltonian.

The Hamiltonian \eqref{eq:F_phi} appears in many different contexts. In statistical physics, the two-dimensional version of the model, which is the one we investigate here, is known to exhibit a Kosterlitz-Thouless transition.\cite{kosterlitz1973ordering} It describes a system of $N$ classical planar spin variables where each spin is coupled to its nearest neighbors on the lattice. It is used to model low-temperature superconductivity, superfluid helium, hexatic liquid crystals, and other phenomena. In the context of quantum field theory, $H$ corresponds to the lattice Landau gauge functional for a compact $U(1)$ lattice gauge theory.\cite{Maas:2011se,Mehta:2009} Each of the stationary points corresponds to a fixed gauge, and a number of interesting physical phenomena, such as the Gribov problem and the Neuberger problem, are related to the stationary points and their properties.\cite{Mehta:2010pe} Furthermore, the Hamiltonian $H$ describes the nearest-neighbor Kuramoto model with homogeneous frequencies.\cite{dorfler2012synchronization} The stationary points of $H$ are the special points in the phase space from the non-linear dynamical systems point of view.\cite{acebron2005kuramoto} Knowing the behavior of the model near the stationary points can greatly enhance our understanding of the full dynamical system.

In an earlier paper on the stationary points of the two-dimensional $XY$ model,\cite{Nerattini:2012pi} specific classes were investigated, predominantly by analytic means. This study was then complemented by a numerical analysis,  focusing on minima and the pathways between them, which are mediated by transition states (stationary points of index one, i.e., with a single negative eigenvalue of the Hessian matrix at the stationary point).\cite{Mehta:2013iea} In the present paper, we compute and analyze general stationary points, without any restrictions on their indices.

The paper is organized as follows: in Sec.\ \ref{s:review} we review previous results for the energy landscapes of $XY$ models. We then describe in Sec.\ \ref{s:methods} the numerical methods employed in the present paper. The numerical results are presented in Sec.\ \ref{s:results}, and our conclusions are summarized in Sec.\ \ref{s:conclusions}.

\section{Previous results}
\label{s:review}

The stationary points of the Hamiltonian \eqref{eq:F_phi} are defined as the solutions of the set of equations
\begin{equation}\label{eq:d_dim_eq}
\frac{\partial H}{\partial\theta_{k}}=\sum_{j\in\mathcal{N}(k)} \sin \left(\theta_j - \theta_k \right) = 0,
\end{equation}
simultaneously for all $k\in\Lambda$. We have performed numerical calculations for periodic boundary conditions as well as for anti-periodic ones. While the choice of boundary conditions affects the stationary points, the qualitative features turned out to be very similar, leading to identical conclusions. For this reason we report here only the results for periodic boundary conditions. Periodic boundary conditions preserve the global $O(2)$ symmetry of the Hamiltonian \eqref{eq:F_phi}. This symmetry implies that all solutions of the stationary point equations \eqref{eq:d_dim_eq} occur in one-parameter families. Continuous families of solutions are harder to deal with numerically, but we avoid this complication by setting the variable $\theta_N$ to zero, thereby explicitly breaking the global $O(2)$ symmetry. Once this symmetry has been broken, the Hamiltonian \eqref{eq:F_phi} has a unique ground state (global minimum) at $\mvec{\theta}\equiv(\theta_1,\dotsc,\theta_N)=(0,\dotsc,0)$, with vanishing energy $H(0,\dotsc,0)=0$.

An analytic study of stationary points was reported in Ref.\ \onlinecite{Casetti:June2003:0022-4715:1091} for the $XY$ model on a fully-connected lattice, i.e., a lattice where every site is considered neighboring to every other site. With such ``mean-field-type'' interactions, exponentially many (in $N$) isolated stationary solutions were found, and also a family of continuous solutions at the maximum value of the energy, even after breaking the global $O(2)$ symmetry. The $XY$ model on a fully-connected lattice is also known as the Kuramoto model in complex systems applications. In Ref.\ \onlinecite{strogatz1991stability} the continuous family of solutions, termed an {\em incoherent manifold}, was observed and discussed.

The stationary points of the one-dimensional $XY$ model with periodic boundary conditions were also studied in Ref.\ \onlinecite{Casetti:June2003:0022-4715:1091}, and a class of stationary points was identified analytically. Subsequently, analytic expressions for all stationary points of that model were reported in Refs.\ \onlinecite{Mehta:2009,Mehta:2010pe}. As in the fully-connected model, some of the solutions were found to be singular and occur in continuous families, even after breaking the global $O(2)$ symmetry. In Refs.\ \onlinecite{Mehta:2009,vonSmekal:2007ns,vonSmekal:2008es} all the stationary points for the one-dimensional model with anti-periodic boundary conditions were characterized. Some analytic results for a one-dimensional $XY$ chain with long-range interactions were reported in Ref.\ \onlinecite{Kastner11}.

A general solution to the stationary equations for the $XY$ model on a cubic lattice in two or higher dimensions turns out to be a formidable task. Constructing certain special classes of analytical solutions is, however, feasible.\cite{Nerattini:2012pi} While most of these special solutions are isolated and nonsingular, singular solutions also exist, either as isolated singular solutions, or as continuous families (even after breaking the global $O(2)$ symmetry of a lattice with periodic boundary conditions). Further progress was made on the numerical side. A crucial step was the observation that the stationary point equations \eqref{eq:d_dim_eq} for the $XY$ model, despite the presence of trigonometric terms, can be viewed as a system of coupled polynomial equations.\cite{Mehta:2009} Polynomial equations are more amenable to numerical techniques such as the polynomial homotopy continuation method,\cite{Mehta:2011xs} a method that has been applied to compute the stationary points of a variety of models in statistical mechanics and particle physics.\cite{Mehta:2011wj,Maniatis:2012ex,Kastner:2011zz,Mehta:2012wk,Mehta:2012qr,Greene:2013ida,Mehta:2013fza,MartinezPedrera:2012rs,He:2013yk} By applying this method to the polynomial form of the $XY$ model, numerical results for the stationary points of the two-dimensional $XY$ model were reported in Refs.\ \onlinecite{Mehta:2009zv,Hughes:2012hg} for small lattices of $3\times 3$ sites.

Other numerical methods have also been applied to the two-dimensional $XY$ model, but they typically find only some of the stationary points or minima,\cite{Hughes:2012hg,Mehta:2014jla} not all of them. Based on data obtained by a conjugate gradient method, it was conjectured in Ref.\ \onlinecite{Hughes:2012hg} that the number of local minima of the two-dimensional $XY$ model increases exponentially with the system size $N$, as expected.\cite{stillingerw84,2003JChPh.11912409W} In a more general XY model, it was shown that the number of minima of the random phase XY model increases exponentially in 2, 3 and  4 dimensions.\cite{Mehta:2014jla} 

The above mentioned minimization methods have a common shortcoming in that they are restricted to relatively small systems of a few tens of lattice sites. In the present paper we push this boundary by about an order of magnitude, treating two-dimensional $XY$ models with up to a hundred lattice sites by means of the numerical techniques introduced in the next section.

\section{Numerical methods}
\label{s:methods}


We used the OPTIM program\cite{optim}  to find minima and transition states for the 2D $XY$ model. 
In particular, we refined $500000$ random initial guesses for all lattice sizes up to $L=10$, i.e.~a total of
$100$ spins. For each solution, $\mvec{\theta}=(\theta_1,\dotsc,\theta_N)$, we 
then considered $\mvec{\theta} \rightarrow -\mvec{\theta}$ and $\mvec{\theta} \rightarrow \mvec{\theta} \pm (\pi, \pi, \dots, \pi)$, i.e., 
the symmetry-related solutions that preserve the index of the second derivative
matrix (Hessian), defined as the number of negative eigenvalues.
Local minima have no negative eigenvalues, while transition states are
here defined according to the geometrical definition, as stationary points
(vanishing gradient)
with precisely one negative eigenvalue.\cite{murrelll68}
OPTIM includes a wide variety of methods for locating stationary points 
of different Hessian index, as well as techniques for characterizing pathways.
A modified version of the limited-memory
Broyden--Fletcher--Goldfarb--Shanno (LBFGS) algorithm \cite{Nocedal80,lbfgs}
was employed for all the minimizations in the present work, since
this approach has proved to be the most efficient in recent benchmarks.\cite{AsenjoSWF13}
\mbox{OPTIM} implements both single- and double-ended \cite{TrygubenkoW04} transition state searches via
either gradient-only or second derivative-based eigenvector-following \cite{Wales92,Wales93d}
and hybrid eigenvector-following algorithms.\cite{munrow99,kumedamw01} Single-ended gradient only methods were generally used here.

In a corresponding paper, we also {\it certify} the numerical solutions we find in the present work using Smale's $\alpha$-theory which can prove if a numerical solution is in the quadratic convergence region of an actual solution of the system.\cite{Mehta:2013zia,Mehta:2014CertifyLong}

\section{Numerical results}
\label{s:results}

\subsection{Numbers of stationary points}
\begin{figure}\centering
\psfrag{N}[tc][tc]{$N$}
\psfrag{ 1}[cr][cr]{$1$}
\psfrag{ 10}[cr][cr]{$10$} 
\psfrag{ 100}[cr][cr]{$100$}
\psfrag{ 1000}[cr][cr]{$1000$}
\psfrag{ 10000}[cr][cr]{$10000$}
\psfrag{ 100000}[cr][cr]{$100000$}
\psfrag{10x}[tc][tc]{$10$}
\psfrag{20x}[tc][tc]{$20$}
\psfrag{30x}[tc][tc]{$30$}
\psfrag{40x}[tc][tc]{$40$}
\psfrag{50x}[tc][tc]{$50$}
\psfrag{60x}[tc][tc]{$60$}
\psfrag{70x}[tc][tc]{$70$}
\psfrag{80x}[tc][tc]{$80$}
\psfrag{90x}[tc][tc]{$90$}
\psfrag{100x}[tc][tc]{$100$}
\psfrag{nspPBC  }[cr][cr]{$n_{\text{sp}}$}
\psfrag{nminPBC  }[cr][cr]{$n_0$}
\psfrag{ntsPBC  }[cr][cr]{$n_1$}
\includegraphics[width=0.90\linewidth]{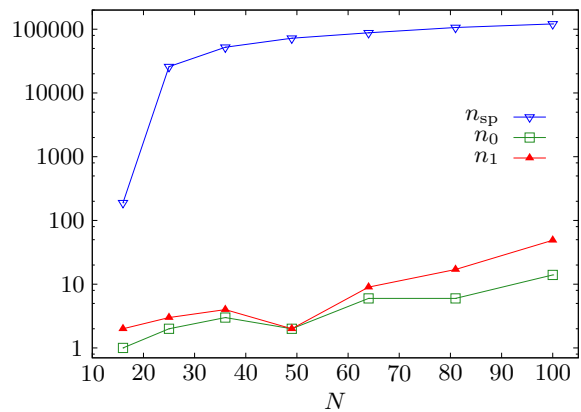}
\caption{\label{fig:plot_N_vs_no_of_SPs}
The total number of stationary points $n_\text{sp}$, the number of minima $n_0$, and the number of transition states $n_1$, as a function of system size $N$.
}
\end{figure}

The number of stationary points of the potential energy function is relevant for a number of applications, 
for example when analyzing the complexity of (spin) glasses. For a
generic potential energy function of a system of $N$ degrees of
freedom, the number of stationary points is expected to grow
exponentially with $N$.\cite{stillingerw84,2003JChPh.11912409W} Hence
a numerical computation will yield only a small subset of all the
stationary points already for moderately large systems. The total
number of stationary points obtained in our numerical calculations
(up to $5\times10^6$ for $N=100$) reflects the computational
effort involved in this study. In spite of this effort, our sample does not
reproduce the actual number of stationary points that the system has (Fig.~\ref{fig:plot_N_vs_no_of_SPs}). 
The situation is different when constraining the search to minima or transition states (stationary points of index 
one).\cite{murrelll68} Their populations, $n_0$ and $n_1$, while also
expected to grow exponentially, are much smaller, and we can expect to
find at least a large fraction of them. This expectation is consistent
with the data in Fig.~\ref{fig:plot_N_vs_no_of_SPs}, where an
exponential increase with $N$ is found for both $n_0$ and $n_1$.

\begin{figure}\centering
\psfrag{1/N}[tc][tc]{$1/N$}
\psfrag{ 0}[cr][cr][1]{$0$}
\psfrag{ 2}[cr][cr]{$2$}
\psfrag{ 4}[cr][cr]{$4$}
\psfrag{ 6}[cr][cr]{$6$}
\psfrag{ 8}[cr][cr]{$8$}
\psfrag{ 10}[cr][cr]{$10$}
\psfrag{ 12}[cr][cr]{$12$}
\psfrag{ 14}[cr][cr]{$14$}
\psfrag{ 16}[cr][cr]{$16$}
\psfrag{ 18}[cr][cr]{$$}
\psfrag{ 0.005}[tc][tc]{$0.005$}
\psfrag{ 0.01}[tc][tc]{$0.010$}
\psfrag{ 0.015}[tc][tc]{$0.015$}
\psfrag{ 0.02}[tc][tc]{$0.020$}
\psfrag{ 0.025}[tc][tc]{$0.025$}
\psfrag{ 0.03}[tc][tc]{$0.030$}
\psfrag{ 0.035}[tc][tc]{$0.035$}
\psfrag{ 0.04}[tc][tc]{$0.040$}
\psfrag{R10  }[cr][cr]{$R_{1,0}~$}
\psfrag{Rsp0   }[cr][cr]{$R_{{\text{sp}},0}$}
\includegraphics[width=0.90\linewidth]{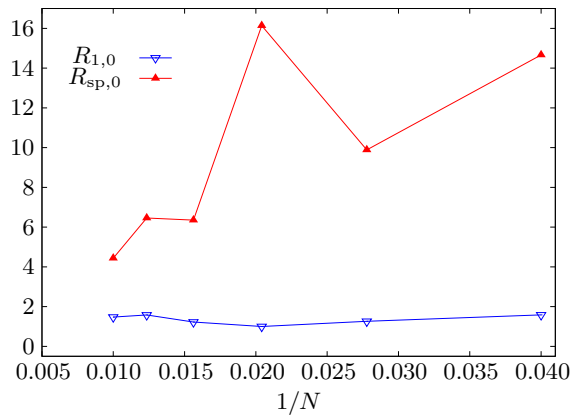}
\caption{\label{TSminvsN_ratio}
Ratio of the logarithm of the number of transition states to the
logarithm of the number of minima {\em vs}.\ $1/N$.
}
\end{figure}

Another way to look at these exponential increases is by considering the ratio of the logarithms
\begin{equation}\label{e:ratio}
R_{I,J}=\frac{\ln n_I}{\ln n_I},
\end{equation}
where $n_I$ and $n_J$ denote the numbers of stationary points with index $I$ and $J$, respectively. 
If $n_I$ and $n_J$ indeed increase exponentially with $N$, $n_I\propto\exp(a_I N)$, 
the ratio of logarithms will be a constant, $R\sim a_I/a_I$, asymptotically for large $N$. 
The same argument also holds for the ratio $R_{\text{sp},J}$, where $n_I$ in the ratio \eqref{e:ratio} 
is replaced by the total number of stationary points $n_\text{sp}=\sum_I n_I$. 
On the basis of our numerical results, we plotted in Fig.\ \ref{TSminvsN_ratio} the ratios 
$R_{1,0}$ and $R_{\text{sp},0}$ {\em vs}.\ the inverse system size $1/N$. The flat, almost-constant behavior of 
$R_{1,0}$ is as expected from the above reasoning and previous theory.\cite{stillingerw84,2003JChPh.11912409W} 
The strong decrease (with increasing $N$) of $R_{\text{sp},0}$ is due to the numerical limitations, indicating that only a small fraction of all stationary points were found.

\begin{figure}\centering
\psfrag{i}[tc][tc]{$i$}
\psfrag{nsp}[cr][cr]{$n_{{\text{sp}}}$}
\psfrag{500}[cr][cr]{$500$}
\psfrag{1000}[cr][cr]{$1000$}
\psfrag{1500}[cr][cr]{$1500$}
\psfrag{2000}[cr][cr]{$2000$}
\psfrag{2500}[cr][cr]{$2500$}
\psfrag{3000}[cr][cr]{$$}
\psfrag{ 0}[r][bl]{$0$}
\psfrag{ 0.2}[tc][tc]{$0.2$}
\psfrag{ 0.4}[tc][tc]{$0.4$}
\psfrag{ 0.6}[tc][tc]{$0.6$}
\psfrag{ 0.8}[tc][tc]{$0.8$}
\psfrag{ 1}[tc][tc]{$1.0$}
\psfrag{L= 4}[cr][cr]{$L=4$}
\psfrag{L= 5}[cr][cr]{$L=5$}
\psfrag{L= 6}[cr][cr]{$L=6$}
\psfrag{L= 7}[cr][cr]{$L=7$}
\psfrag{L= 8}[cr][cr]{$L=8$}
\psfrag{L= 9}[cr][cr]{$L=9$}
\psfrag{L= 10}[cr][cr]{$L=10$}
\includegraphics[width=0.97\linewidth]{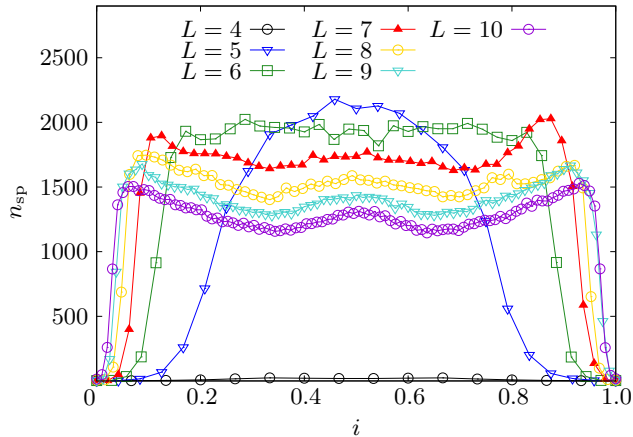}
\caption{\label{fig:N_saddlevsIdens}
The number $n_{\text{sp}}$ of stationary points as a function of the
index density $i$, shown for various lattice sizes $N=L\times L$. 
}
\end{figure}

A more detailed analysis of the index-dependence of the numbers of stationary points is shown in Fig.\ \ref{fig:N_saddlevsIdens}. In this plot the numbers of stationary points of a given index $I$ are shown {\em vs}.\ the index density $I/(N-1)$. The observed behavior is in part due to the properties of the system, and in part determined by the finite computational resources. The steep increase or decrease at the flanks of the curves (i.e., around $i=0$ and $i=1$) reflects the actual behavior of the total number of stationary points of that index, which is expected to follow a binomial distribution.\cite{2003JChPh.11912409W} The flat region inbetween (except for the $4\times4$ and $5 \times 5$ lattices) is an artefact of the numerical limitations.

\subsection{Energies at stationary points}
\label{s:energies}
A physical system at a given energy (or temperature) will sample 
a subset of the energy landscape. 
It is therefore not surprising that the stationary energies, i.e., the Hamiltonian \eqref{eq:F_phi} evaluated at the various stationary points, play an important role in energy landscape applications. 

\begin{figure}\centering
\includegraphics[width=1.0\linewidth]{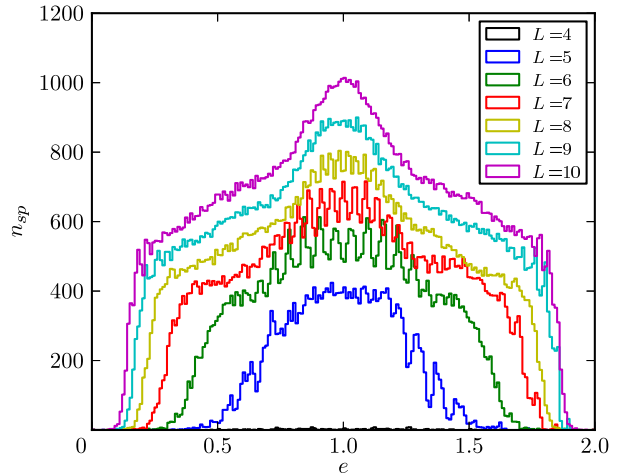}
\caption{\label{f:histE_PBC}
The number $n_\text{sp}$ of stationary points in intervals $[e,e+0.01]$ of the energy density $e=E/N$ with $N=L\times L$. }
\end{figure}

\begin{figure}\centering
\includegraphics[width=0.93\linewidth]{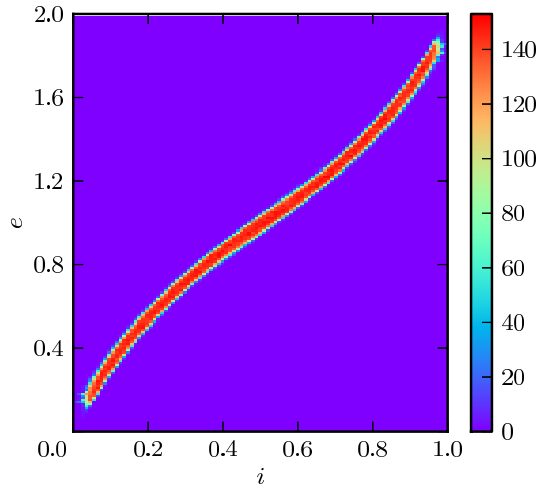}
\caption{\label{fig:i_vs_e_of_I} 
Density plot of the frequencies of the occurrence of stationary points of a certain index density $i$ and energy density $e$. The frequencies are obtained as the number of stationary points with $e\in\big[10^{-2}n,10^{-2}(n+1)\big)$ and $i\in\big[10^{-2}m,10^{-2}(m+1)\big)$, where $n,m\in\{0,1,2,\dotsc\}$. The distribution is sharply peaked around a bent curve in the $(i,e)$-plane, indicating the strong correlation between index and energy densities. The plot shown is for a lattice of size $10 \times 10$. The distributions for smaller lattices look similar, but are less sharply peaked.
}
\end{figure}

Analyzing the number of stationary points as a function of energy, we
find the bell-shaped distribution shown in Fig.\ \ref{f:histE_PBC}. As
for the number of stationary points as a function of the index density
in Fig.\ \ref{fig:N_saddlevsIdens}, the behavior reflects in
part the properties of the system and in part the numerical
limitations. The two plots are in fact closely linked, as energy and
index density are strongly correlated, as illustrated in Fig.\
\ref{fig:i_vs_e_of_I}. Such a correlation is expected: The
minima (stationary points of index 0) will typically be of lower
energy than the maxima (stationary points of index $N$). Or, more
generally, the energy of stationary points of index $I+1$ is expected
to be higher than for those of index $I$.\cite{murrelll68} Based on
this observation we conclude that, similar to Fig.\
\ref{fig:N_saddlevsIdens}, the steep flanks of the curves in Fig.\
\ref{f:histE_PBC} reflect the actual dependence of the number of
stationary points on the energy, whereas the flatter regions of the plot correspond to energies where the actual numbers of stationary points are so large that only a small fraction is found numerically.

\subsection{Energy Differences}
\label{s:differences}
The difference in energy between two stationary points can determine thermodynamic and dynamic properties. 
For example, energy barriers appear exponentially in unimolecular rate theory in the canonical ensemble.\cite{Forst73}
Here, instead of looking at energy differences between specific states, we follow a statistical approach, investigating the frequency of occurrence of energy gaps of a certain size. Somewhat in the spirit of Wigner's level statistics,\cite{MehtaBook} we focus on the differences
\begin{equation}
\Delta_i=e_i-e_{i-1}
\end{equation}
between neighboring values of the stationary energy densities $e_i=H(\mvec{\theta}_i^\text{s})/N$. The various stationary points $\theta_i^\text{s}$ are sorted such that the energy densities $e_i$ form an increasing sequence. On the basis of the differences $\Delta_i$ between neighboring stationary energies, all other differences can be computed.

\begin{figure}\centering
\includegraphics[width=0.97\linewidth]{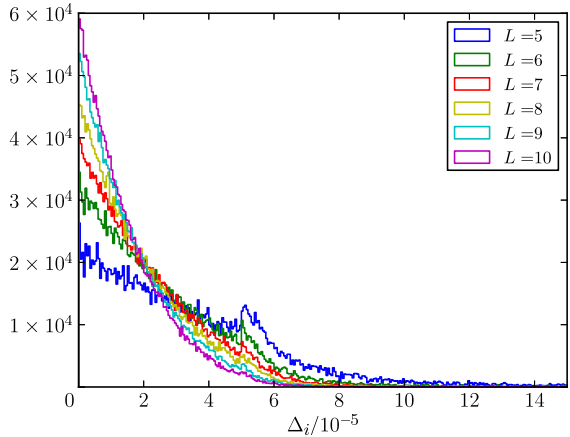}
\caption{\label{histEdiff_4_linear}
Relative frequency of the occurrence of stationary energy differences
$\Delta_i$, obtained as the number of $\Delta_i$-values in the binning
intervals $\big[5 \times 10^{-7}n,5 \times 10^{-7}(n+1)\big)$ with
$n\in\{0,1,2,\dotsc\}$. 
}
\end{figure}

In Fig.\ \ref{histEdiff_4_linear}, the relative frequency for the occurrence of energy differences $\Delta_i$ is shown for various system sizes $N$. For all values of $N$, the maximum relative frequency is attained for the smallest binning interval, $\Delta_i\in[0,0.05)$. The overall trend of all the curves is a monotonic decrease for larger $\Delta_i$, superimposed by fluctuations. At least for the smaller system sizes shown, the relative frequency of small $\Delta_i$ values grows with increasing system size. 
Such behavior is expected: An exponentially (in $N$) growing number of stationary energies has to be accommodated in a finite interval of energy densities, and this 
observation implies that typical distances between neighboring energy densities will decrease dramatically. 
For the largest system sizes studied ($9\times9$ and $10\times10$) the tendency towards smaller $\Delta_i$ is virtually absent, 
which we attribute to the fact that only a small fraction of the exponentially many stationary points could be computed for these system sizes, with absolute sample sizes that are virtually $N$-independent.

\begin{figure}\centering
\includegraphics[width=0.9\linewidth]{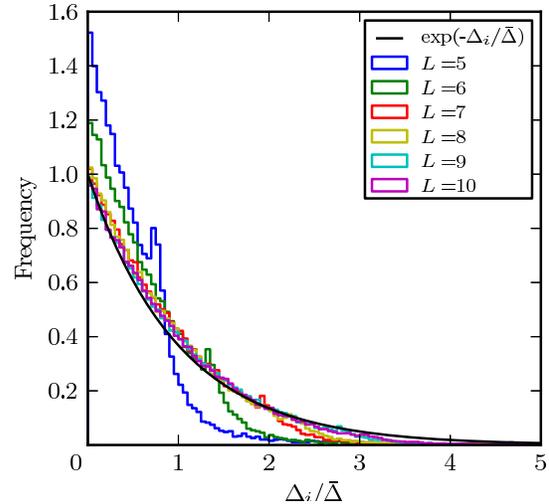}
\caption{\label{histEdiff_11}
As in Fig.\ \ref{histEdiff_4_linear}, but for the normalized energy
differences $\Delta_i/\overbar{\Delta}$ and binning intervals
$\big[0.05n,0.05(n+1)\big)$ with $n\in\{0,1,2,\dotsc\}$. 
The collapsed data nicely follow a decaying exponential $\exp(-\Delta_i/\overbar{\Delta})$.
}
\end{figure}

The trivial tendency towards smaller $\Delta_i$-values, caused by the increasing number of stationary points, can be eliminated by normalizing the $\Delta_i$ to 
a unit average. This normalization is achieved by computing the sample average
\begin{equation}
\overbar{\Delta}=\frac{1}{M}\sum_{i=1}^M\Delta_i,
\end{equation}
where $M$ is the sample size. The normalized energy differences $\Delta_i/\overbar{\Delta}$ are shown in Fig.\ \ref{histEdiff_11} for various system sizes $N$. With the exception of the very small lattice sizes of $5\times5$ and $6\times6$, the various curves now collapse onto each other, indicating that the distribution of normalized energy differences is largely independent of the system size, and presumably converges in the large-$N$ limit. The collapsed data appear to follow a decaying exponential $\exp(-\Delta_i/\overbar{\Delta})$; note that no fitting parameter is involved. The use of such a decaying exponential is inspired by Wigner's level statistics for the differences between neighboring energy eigenvalues of the Hamiltonian of an integrable quantum mechanical system.

\subsection{Hessian determinant at stationary points}
\label{s:Hesse}
The energies at stationary points, discussed in Secs.~\ref{s:energies} and \ref{s:differences}, give the leading, zeroth order contribution of a Taylor expansion around a stationary point. The next nonvanishing term is quadratic, with the expansion coefficients given by the elements of the Hessian matrix. 
The quadratic expansion corresponds to standard normal mode analysis and generates the harmonic vibrational density of states, which can be employed to analyze equilibrium thermodynamic properties, as well as rate coefficients.\cite{Wales03}

One way to condense the information contained in the many matrix elements of the Hessian matrix into a single number is by computing its index $I$, 
as introduced in Sec.\ \ref{s:methods}, where only the {\em signs}\/ of the eigenvalues enter. To condense information about the {\em magnitude}\/ of the eigenvalues into a single number, we compute the determinant at a stationary point (equal to the product of all the eigenvalues). Zero eigenvalues that result from translational or rotational symmetry must first be eliminated from consideration, either by projection, shifting, or coordinate transformation.\cite{Wales03} Roughly speaking, a small value of the determinant corresponds to a ``flatter'' stationary point, and a large value to a ``narrower'' one, with lower vibrational entropy. The Hessian determinant at a stationary point $\mvec{\theta}^\text{s}$, and more precisely its rescaled version
\begin{equation}
D=\left|\det\mathcal{H}(\mvec{\theta}^\text{s})\right|^{1/N},
\end{equation}
has been proposed as an indicator for (the absence of) phase transitions in the limit of large system size; see Refs.\ \onlinecite{kastner2007phase,kastner2008phase,KSS08} for details.

\begin{figure}\centering
\psfrag{e}[tc][tc]{$e$}
\psfrag{D}[tc][tc]{$D$}
\psfrag{0x}[r][bl]{$0$}
\psfrag{0.5x}[tc][tc]{$0.5$}
\psfrag{1.0x}[tc][tc]{$1.0$}
\psfrag{1.5x}[tc][tc]{$1.5$}
\psfrag{2.0x}[tc][tc]{$2.0$}
\psfrag{0.5y}[cr][cr]{$0.5$}
\psfrag{1.0y}[cr][cr]{$1.0$}
\psfrag{1.5y}[cr][cr]{$1.5$}
\psfrag{2.0y}[cr][cr]{$2.0$}
\psfrag{L = 4}[cr][cr]{$L=~4$}
\psfrag{L = 5}[cr][cr]{$L=~5$}
\psfrag{L = 6}[cr][cr]{$L=~6$}
\psfrag{L = 7}[cr][cr]{$L=~7$}
\psfrag{L = 8}[cr][cr]{$L=~8$}
\psfrag{L = 9}[cr][cr]{$L=~9$}
\psfrag{L = 10}[cr][cr]{$L=10$}
\includegraphics[width=0.93\linewidth]{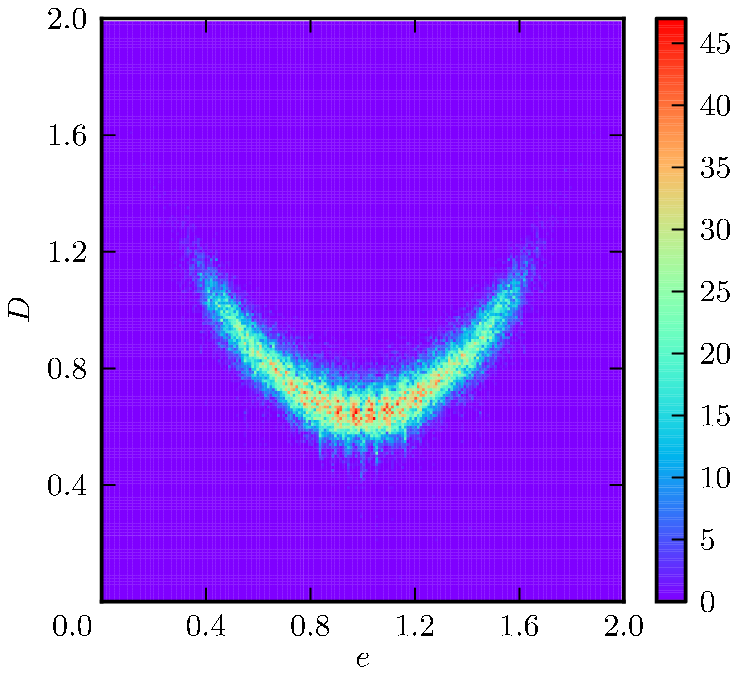}
\includegraphics[width=0.93\linewidth]{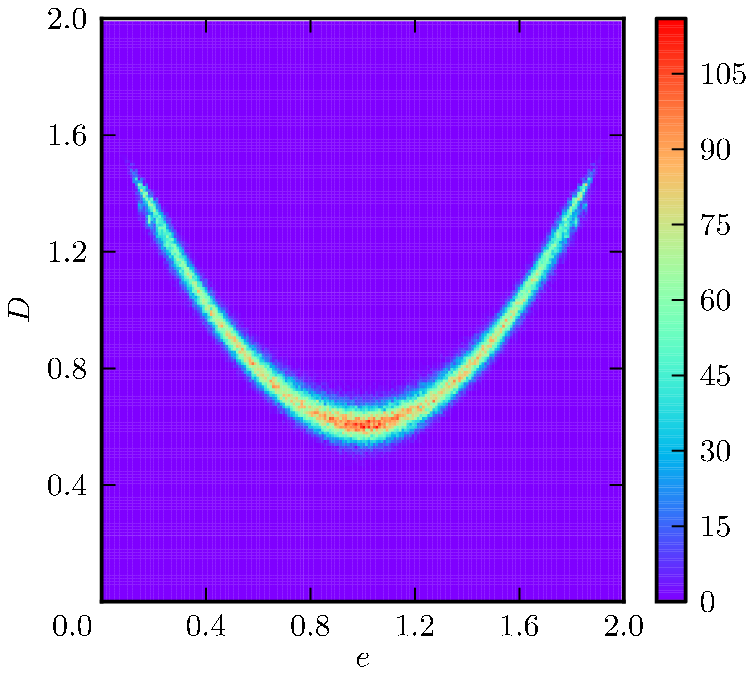}
\caption{\label{e_vs_det_H}
Density plots of the frequencies of the occurrence of stationary points of energy density $e$ and rescaled determinant $D$. The frequencies are obtained as the number of stationary points with $e\in\big[10^{-2}n,10^{-2}(n+1)\big)$ and $D\in\big[10^{-2}m,10^{-2}(m+1)\big)$, where $n,m\in\{0,1,2,\dotsc\}$. 
The plots are for lattice sizes $6\times6$ (top) and $10 \times 10$ (bottom).
}
\end{figure}

For each stationary point computed, the pair $(e,D)$ is calculated, where $e=H(\mvec{\theta}^\text{s})/N$ is the energy density at the stationary point. The density plots in Fig.\ \ref{e_vs_det_H} illustrate that the rescaled determinant $D$ and the energy density $e$ are strongly correlated, accumulating around a bow-shaped curve in the $(e,D)$-plane. With increasing system size, the distribution becomes more sharply peaked around this curve. This observation suggests that, in the limit of infinite system size, the rescaled Hessian determinant $D$ is sharply localized for each value of $e$, behaving like a thermodynamic quantity.

\subsection{Eigenvalues}
\label{s:eigenvalues}
Various quantities have been studied previously in relation to the Hessian eigenvalues.\cite{DoyeW02} 
In the harmonic normal mode approximation, the vibrational partition function and 
associated density of states are determined by the product of normal mode frequencies.
The corresponding transition state theory \cite{Forst73} rate constants also depend
on these frequencies, which are obtained from the mass-weighted Hessian eigenvalues.\cite{Wales03}
Some interesting properties have also been examined for 
the smallest Hessian eigenvalue in terms of catastrophe theory.\cite{Wales01} 

Let $\lambda^{(I)}_i$ denote the lowest eigenvalue of a stationary point of index $I$. We can average over the lowest eigenvalue at each stationary point for a particular index $I$.
\begin{align}
\langle \lambda \rangle^{(I)} & = \frac{1}{n_\text{sp}(I)} \sum_{i=1}^{n_\text{sp}(I)} \lambda^{(I)}_i \label{avg_eig}
\end{align}

\begin{figure}\centering
\includegraphics[width=0.9\linewidth]{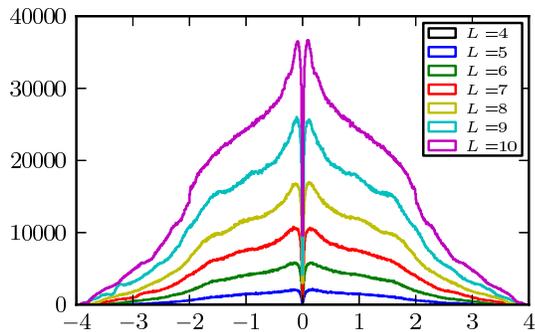}
\caption{Frequency of occurrence of eigenvalues of the Hessian in the binning
intervals $\big[0.01n,0.01(n+1)\big)$ with
$n\in\{0,1,2,\dotsc\}$.
}
\label{fig:hist_evalues}
\end{figure}

In Fig.~\ref{fig:hist_evalues}, we plot a histogram of all the eigenvalues of the Hessian matrices computed at all the stationary points we obtained.
The plots seem to become bell-shaped curves as $N$ increases, with a sharp discontinuity at the origin representing the fact that we have only considered nonsingular stationary solutions in this study.

\begin{figure}[h]\centering
\psfrag{i}[tc][tc]{$i$}
\psfrag{0.0x}[tc][tc]{$~~~~0.0$}
\psfrag{0.2x}[tc][tc]{$0.2$} 
\psfrag{0.4x}[tc][tc]{$0.4$}
\psfrag{0.6x}[tc][tc]{$0.6$}
\psfrag{0.8x}[tc][tc]{$0.8$}
\psfrag{1.0x}[tc][tc]{$1.0$}
\psfrag{-4}[cr][cr]{$-4$}
\psfrag{-3}[cr][cr]{$-3$}
\psfrag{-2}[cr][cr]{$-2$}
\psfrag{-1}[cr][cr]{$-1$}
\psfrag{ 0}[cr][cr]{$0$}
\psfrag{ 1}[cr][cr]{$1$}
\psfrag{loweig}[cr][cr]{$~~\langle \lambda \rangle ^{(I)}$}
\psfrag{L= 4}[cr][cr]{$L=4$}
\psfrag{L= 5}[cr][cr]{$L=5$}
\psfrag{L= 6}[cr][cr]{$L=6$}
\psfrag{L= 7}[cr][cr]{$L=7$}
\psfrag{L= 8}[cr][cr]{$L=8$}
\psfrag{L= 9}[cr][cr]{$L=9$}
\psfrag{L= 10}[cr][cr]{$L=10$}
\includegraphics[width=0.90\linewidth]{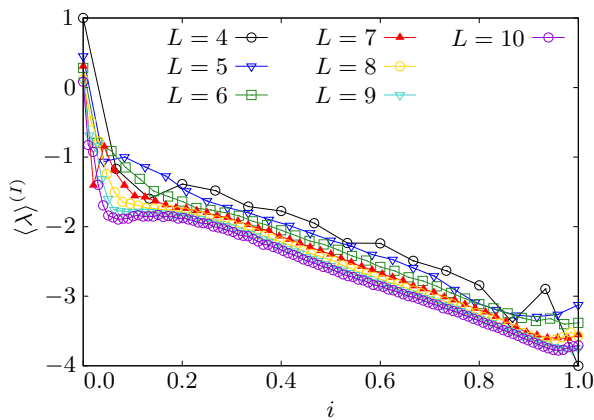}
\caption{The average lowest eigenvalue $\langle \lambda \rangle^{(I)}$
  as a function of  index density $i$.
}
\label{fig:av_lowest_evalue}
\end{figure}

In the binary Lennard-Jones liquid at constant volume, a linear decrease of the average of the lowest eigenvalues of the Hessian is seen when the energy is increased
above the threshold energy at which the first stationary points with higher index are found,\cite{PhysRevLett.85.5360}
i.e., a linear decrease with $i$. 
In atomic clusters bound by the pairwise Lennard-Jones potential,
\cite{jonesi25} the behavior of the average lowest eigenvalue was shown to tend to have a quadratic dependence on $i$ as the number of particles 
increased.\cite{DoyeW02}
In the present work, we observe a linear decrease of the lowest eigenvalue as a function of $i$ beyond 
a threshold value for $i$ in Fig.~\ref{fig:av_lowest_evalue}.
This behavior is therefore closer to the bulk structural glass former than to an atomic cluster.

\section{Discussion and Conclusions}
\label{s:conclusions}

We have numerically computed stationary points of the potential energy landscape of the two-dimensional $XY$ model on a square lattice for systems of up to $N=10\times10$ sites. Since the number of stationary points is believed to grow exponentially with $N$, we can in general sample only a small fraction of them. As a consequence, the results reflect properties of the underlying energy landscape, but also of the restricted sample size. The main motivation for the present study was to better understand the interplay of physical features and the restricted sample size, as this is an important aspect in the application of numerical techniques to the study of energy landscapes of large systems. 

The interplay of physical features and the restricted sample size becomes particularly obvious, and can be analyzed by classifying the stationary points by their Hessian index $I$. Stationary points of indices around $I=N/2$ are much more numerous than those of indices close to $0$ or close to $N$. For this reason, the available sample sizes faithfully reproduce the physical properties of stationary points of small or large indices, while the numerical limitations become dominant for intermediate values of $I$. These different regimes, and the crossover between them, are illustrated from various perspectives in Figs.\ \ref{fig:plot_N_vs_no_of_SPs}--\ref{f:histE_PBC}. In the regime of small or large indices where the sample sizes are sufficient, exponentially increasing numbers of stationary points, a binomial distribution in index density, and other properties expected from approximate theoretical arguments are nicely confirmed.

Restricted sample sizes pose a problem for quantities that are---like the above examples---based on the numbers of stationary points. In Secs.\ \ref{s:differences}--\ref{s:eigenvalues} we have studied several other properties of the energy landscape where the problem of restricted sample size can be avoided, or at least attenuated. Examples include the (rescaled) determinants of Hessian matrices at stationary points in Sec.\ \ref{s:Hesse} and the averaged lowest eigenvalues in Sec.\ \ref{s:eigenvalues}. In Sec.\ \ref{s:differences} we have analyzed the distribution of the distances between neighboring stationary energy levels. While such distributions are frequently studied for eigenenergies in the context of quantum chaos, their application in the context of energy landscapes is novel. The Poisson-type distributions we find are familiar from the quantum mechanical counterpart and they seem to be little affected by the small sample size of the numerical calculations.

\begin{acknowledgments}
D.\,M.\ was supported by a DARPA Young Investigator Award and by the ERC. 
C.\,H.\ acknowledges support from the Science and Technology Facilities Council and the Cambridge Home and European Scholarship Scheme. 
M.\,K.\ acknowledges support by the Incentive Funding for Rated Researchers program of the National Research Foundation of South Africa. 
D.\,J.\,W.\ gratefully acknowledges support from the EPSRC and the ERC.
\end{acknowledgments}

\end{document}